\documentclass[12pt]{article}

\usepackage{graphics}
\usepackage{epsfig}
\input epsf

\title{Beauty production in two-photon interactions at LEP2: \\
 $k_T$-factorization versus data}

\author{A.V.~Lipatov, N.P.~Zotov}

\begin{document}

\maketitle

\begin{center}

{\it D.V. Skobeltsyn Institute of Nuclear Physics,\\ 
M.V. Lomonosov Moscow State University,
\\119991 Moscow, Russia\/}\\[3mm]

\end{center}

\vspace{0.5cm}

\begin{center}

{\bf Abstract }

\end{center}

Inclusive beauty quark production in photon-photon collisions 
at CERN LEP2 is considered in the framework of the $k_T$-factorization 
approach. Both direct and resolved photon contributions are taken 
into account. The unintegrated gluon distributions in 
a photon are either obtained from the full CCFM evolution equation or
from the Kimber-Martin-Ryskin prescription.
The predicted beauty cross section 
reasonably agrees with the recent experimental data taken by the ALEPH
collaboration. We argue that theoretical and experimental studies 
of the azimuthal correlations in heavy quark production at high
energies can serve as a crucial probe of the unintegrated gluon densities.

\vspace{0.8cm}

\noindent
PACS number(s): 12.38.-t, 13.85.-t

\vspace{1.0cm}

The problem  of 
beauty quark production  at high energies continues to be a subject 
of pointed discussions and intense theoretical studies up to now~[1]. 
First results~[2] on the $b$-quark cross section in 
$ep$-collisions at HERA were significantly higher than the QCD predictions
calculated at next-to-leading 
order (NLO) approximation. Similar observations were made in hadron-hadron
collisions at the Fermilab Tevatron~[3] and also in photon-photon 
interactions at LEP2~[4]. In the latter case the 
theoretical NLO QCD predictions  were below the experimental data 
by three standard deviations. Although the latest measurements~[5] do not 
confirm the large excess of the first HERA data over the NLO QCD, the 
problem is not solved so far.
The disagrement between the experimental data at the Tevatron and NLO 
QCD predictions was reduced by adopting a special nonperturbative 
fragmentation 
function of the $b$-quark into the $B$-meson~[6]\footnote{A more exotic
solution to this problem was proposed in~[7].}.

From our point of view, a  more adequate solution  was found~[8] in the 
framework of $k_T$-factorization approach~[9]. 
The $k_T$-factorization approach has also been used  for a detailed 
description of numerous experimental data on $b$-quark 
production at HERA~[10]. However the problem of the $b$-quark
production in $\gamma \gamma$ interactions
is not solved so far in the $k_T$-factorization approach~[11--14].

Recently the ALEPH collaboration at LEP2 has presented the result on
open beauty production cross section in $\gamma \gamma$ 
collisions~[15]. This is the
first published measurement in which the lifetime information has been 
used to identify the heavy flavor  in two-photon 
physics\footnote{The previous 
measurements by L3 and OPAL collaborations~[4] were based on
a fiting  the transverse momentum of leptons 
with respect to jets.}. The cross section of the process $e^+ e^-
\to e^+ e^- b \bar b \, X$
has been found to be $5.4 \pm 0.8 \,{\rm (stat.)} \pm 0.8 
\,{\rm (syst.)}$~pb which is
fully inconsistent with the previous results quoted by the L3 and 
OPAL collaborations~[4],
namely $12.8 \pm 1.7 \,{\rm (stat.)} \pm 2.3 \,{\rm (syst.)}$~pb
and $14.2 \pm 2.5 \,{\rm (stat.)} ^{-4.8}_{+5.3} \,{\rm (syst.)}$~pb,
respectively.
In the present note we would like to demonstrate that the ALEPH experimental 
data can be described in the $k_T$-factorization approach also and
to propose an additional test to distinguish the different 
unintegrated gluon distribution functions, which are the main ungredient 
of the $k_T$-factorization (see, for example,~[16]).  

Theoretically, heavy quarks in $\gamma \gamma$ collisions can be 
produced via direct and resolved production mechanisms. In
the direct events,  two photons couple 
directly to a heavy quark pair. This contribution is governed by simple QED
amplitudes (which are independent of the gluon density in the photon). 
In the resolved events, one photon ("single-resolved") or both 
photons ("double-resolved") fluctuate into a hadronic state and a gluon or 
a quark from
of this hadronic fluctuation takes part in the hard interaction.
At LEP2 conditions the 
heavy quark production via the double resolved processes is highly 
suppressed~[17] and, therefore, it
will not be taken into account in our analysis.

The single-resolved contribution to the $\gamma \gamma \to b\bar b$ process
is dominated by the gluon component of the photon and has the following 
form in the $k_T$-factorization approach:
$$
  { d\sigma_{\rm 1-res} (\gamma \gamma \to b\bar b \, X) \over dy d{\mathbf p}_{T}^2 } = \int {1\over 16\pi (x s)^2 (1 - \alpha)} {\cal A}_{\gamma}(x,{\mathbf k}_T^2,\mu^2) 
  |\bar {\cal M}|^2(\gamma g^* \to b\bar b) d{\mathbf k}_T^2 {d\phi_b \over 2\pi} {d\phi \over 2\pi}, \eqno (1)
$$

\noindent 
where ${\cal A}_{\gamma}(x,{\mathbf k}_{T}^2,\mu^2)$ is the
unintegrated gluon distribution
in the photon, $|\bar {\cal M}|^2(\gamma g^* \to b\bar b)$ is the 
off-shell 
(i.e. depending on the initial gluon virtuality) matrix element squared, 
$s$ is the total c.m. frame energy and 
$\alpha = \sqrt {m_b^2 + {\mathbf p}_{T}^2}\exp(y)/\sqrt s$.
The produced beauty quark has the transverse momentum ${\mathbf p}_{T}$, 
rapidity $y$ and azimuthal angle $\phi_b$.
The initial off-shell gluon has a fraction $x$ of the parent photon's 
longitudinal 
momentum, the non-zero transverse momentum ${\mathbf k}_T$ (${\mathbf k}_T^2 = - k_T^2 \neq 0$) 
and azimuthal angle $\phi$. In accord with the $k_T$-factorization
prescription~[9],
the off-shell gluon spin density matrix is taken in the form
$$
   \epsilon^\mu (k) \epsilon^{*\,\nu} (k) = { k_T^\mu k_T^\nu \over {\mathbf k}_T^2}. \eqno(2)
$$

\noindent 
In all other respects our calculations follow the standard Feynman rules.
The analytic expression for the 
$|\bar {\cal M}|^2 (\gamma g^* \to b\bar b)$ is given in our previous
paper~[13]. Note that if we average Eq.~(1) over the azimuthal angle $\phi$ 
and take the limit ${\mathbf k}_{T}^2 \to 0$, we recover the well-known 
formulas 
corresponding to the leading-order (LO) QCD calculations.

The recent experimental data~[15] refer to beauty quark 
production in the $e^+ e^-$ collisions. In order to obtain
the corresponding cross sections, the $\gamma \gamma$ cross sections need 
to be weighted with the photon flux in the electron:
$$
  d\sigma(e^+ e^- \to e^+ e^- b\bar b \, X) = \int f_{\gamma/e}(x_1)dx_1 \int f_{\gamma/e}(x_2)dx_2\,d\sigma(\gamma \gamma \to b\bar b \,X), \eqno (3)
$$

\noindent
where we use the Weizacker-Williams approximation for the photon
distribution in the electron:
$$
  f_{\gamma/e}(x) = {\alpha_{em} \over 2\pi}\left({1 + (1 - x)^2\over x}\ln{Q^2_{\rm max}\over Q^2_{\rm min}} + 
  2m_e^2 x\left({1\over Q^2_{\rm max}} - {1\over Q^2_{\rm min}} \right)\right). \eqno (4)
$$

\noindent
Here $\alpha_{em}$ is the fine structure constant, $m_e$ is the electron 
mass, $Q^2_{\rm min} = m_e^2x^2/(1 - x)^2$ and $Q^2_{\rm max} = 
6$~GeV$^2$~[15].

The unintegrated gluon distribution in the photon ${\cal 
A}_{\gamma}(x,{\mathbf k}_{T}^2,\mu^2)$
can be obtained from the analytical or numerical solution of the BFKL or CCFM
evolution equations. In order to estimate the degree of theoretical uncertainty
connected with the choice of unintegrated gluon densities,
in the numerical calculations we  tested 
two different sets, namely the CCFM~[12] and KMR~[18] ones. First of them 
was 
obtained in~[12] from the full CCFM equation formulated for the photon, 
and 
the second one was obtained from the usual (collinear) parton 
densities\footnote{In the numerical calculations we have used the 
standard GRV~(LO) parametrizations~[19] of the collinear quark and gluon 
distributions.} using the Kimber-Martin-Ryskin prescription~[18]. These
distributions are widely discussed in the literature (see, for example,~[16]).
Other essential parameters were taken as follows: the $b$-quark mass
$m_b = 4.5 \pm 0.1$~GeV and the renormalization and factorization 
scale $\mu = \xi \sqrt{m_b^2 + \langle {\mathbf p}_{T}^2 \rangle}$, 
where $\langle {\mathbf p}_{T}^2 \rangle$ is set to the average
${\mathbf p}_{T}^2$ of the beauty quark and antiquark.
In order to investigate the scale dependence of our 
results we  vary the scale parameter
$\xi$ between $1/2$ and 2 about the default value $\xi = 1$.
For completeness, we use the LO formula for the coupling constant 
$\alpha_s(\mu^2)$ 
with $n_f = 4$ active quark flavours and $\Lambda_{\rm QCD} = 200$~MeV, 
such that $\alpha_s(M_Z^2) = 0.1232$. The multidimensional integration
has been performed by means of the Monte Carlo technique, using 
the routine \textsc{vegas}~[20].
The full C$++$ code is available from the authors on 
request\footnote{lipatov@theory.sinp.msu.ru}. This code is
identical to that used in~[13, 14].

The results of our calculations are displayed in Figs.~1 --- 4.
Fig.~1 confronts the total cross section $\sigma(e^+ e^- \to e^+ e^- b \bar b \, X)$
calculated as a function of the total c.m. energy $\sqrt s$ with recent experimental 
data~[15] taken by the ALEPH collaboration. The solid and dash-dotted 
lines 
correspond to the results obtained with the CCFM and KMR unintegrated gluon 
densities, respectively. The upper and lower dashed lines correspond to 
the CCFM gluon density with $b$-quark mass and scale variations as it was 
described above. 
Separately shown (as a dotted line) is the contribution from the direct production 
mechanism $\gamma \gamma \to b \bar b$.
It is clear that at $\sqrt s \sim 200$~GeV the cross section is mostly controlled
by the single-resolved contribution, i.e. $\gamma g^* \to b \bar b$ 
subprocess.
Despite the fact that the central predictions are slightly lower than the 
measured cross section, we observe a reasonable agreement between our calculations 
and the ALEPH experimental data~[15] within the theoretical and 
experimental uncertainties.
The CCFM-evolved gluon density gives  slightly larger 
cross section compared to the KMR one, where the small-$x$ logarithms 
are not taken into account~[18].
A similar effect (but much more clear) has been demonstrated in~[10] 
where, in particular, 
the beauty photo- and lepto-production at HERA has been studied.
Note that the sensitivity of our results to the variations in the scale 
$\mu$ and beauty mass $m_b$ is rather large. However, this sensitivity is 
of the same order approximately as in the massive NLO QCD 
calculations~[21].

The transverse momentum and pseudo-rapidity distributions calculated
at the averaged total $e^+ e^-$ energy $\sqrt s = 196$~GeV ($130 < \sqrt s 
< 209$~GeV) 
are shown in Figs.~2 and~3. As a representative example, we have used
the following cuts: $p_T < 20$~GeV and $|\eta| < 2$. In our calculations
we took into account for both the beauty quarks and anti-quarks.
One can see again that the difference between the CCFM and KMR predictions
is not significant, except at large $p_T$ (namely $p_T \sim 10$~GeV) 
only.
A similar observation was also made~[14] in the case of 
charm production at LEP2. It was shown that the shape and the absolute 
normalization of $D^*$ transverse momentum and pseudo-rapidity distributions
practically do not depend on the unintegrated gluon density.

We would like to stress that further understanding of the process 
dynamics may be obtained from the 
angular correlation between the transverse momenta of the produced quarks.
These quantities are particularly sensitive to high-order corrections.
So, in the naive LO collinear approximation of QCD, the distribution 
over $\Delta \phi = \phi_b - \phi_{\bar b}$ must be simply a delta function 
$\delta(\Delta \phi - \pi)$ since the produced quarks are back-to-back in 
the transverse plane. Large deviations from these values may come from 
higher-order QCD effects. In the $k_T$-factorization 
approach, taking into account the non-vanishing initial gluon
transverse momentum ${\mathbf k}_{T}$ leads to 
the violation of this back-to-back kinematics even at leading order.
It is an illustration to the fact that the LO $k_T$-factorization 
formalism incorporates a large part of standard (collinear) high-order 
corrections (see also~[9, 16] for more information).
The differential cross section $d\sigma/d\Delta\phi$
calculated at $\sqrt s = 196$~GeV is shown in Fig.~4.
One can see that the shape of this distribution predicted by the CCFM and 
KMR gluon densities are strongly differ from each other. 
At large $\Delta \phi \sim \pi$ both gluon densities under
consideration give  similar results, whereas at low $\Delta \phi \sim 0$
the difference is about a factor of 2 in the absolute normalization.
This fact is directly
connected with the properties of non-collinear evolution model.
Therefore these correlations can be used  to constraint the
unintegrated gluon distributions. A similar effect
was also pointed out in the case of beauty production at the Tevatron~[8].

In conclusion, we would like to emphasize additionally that the 
$k_T$-factorization approach supplemented with the CCFM-evolved
gluon density agrees well with the numerous data on the $b$-quark 
production at HERA and 
Tevatron (without any special assumption on the $b$-quark to $B$-meson fragmentation function), 
as it was demonstrated earlier in~[10, 8]. 
So we can conclude that at present there is no contradiction between
the CCFM-based theoretical predictions and available data on the beauty production at
high energies, and we believe that the $k_T$-factorization holds a 
possible key to understanding the production dynamics at high energies.

We thank H.~Jung for offering the CCFM code for the 
unintegrated gluon distributions used in our calculations and 
S.P.~Baranov for careful reading of the manuscript.
The authors are very grateful to P.F.~Ermolov for the support and DESY 
Directorate for the support in the framework of Moscow --- DESY project on
Monte-Carlo implementation for HERA --- LHC.
A.V.L. was supported in part by the grant of President of 
Russian Federation (MK-9820.2006.2) and the grant of  Helmholtz --- Russia 
Joint Research Group (HRJRG-002). Also this research was supported by the 
FASI of Russian Federation (grant NS-8122.2006.2).

\newpage

\begin{figure}
\begin{center}
\epsfig{figure=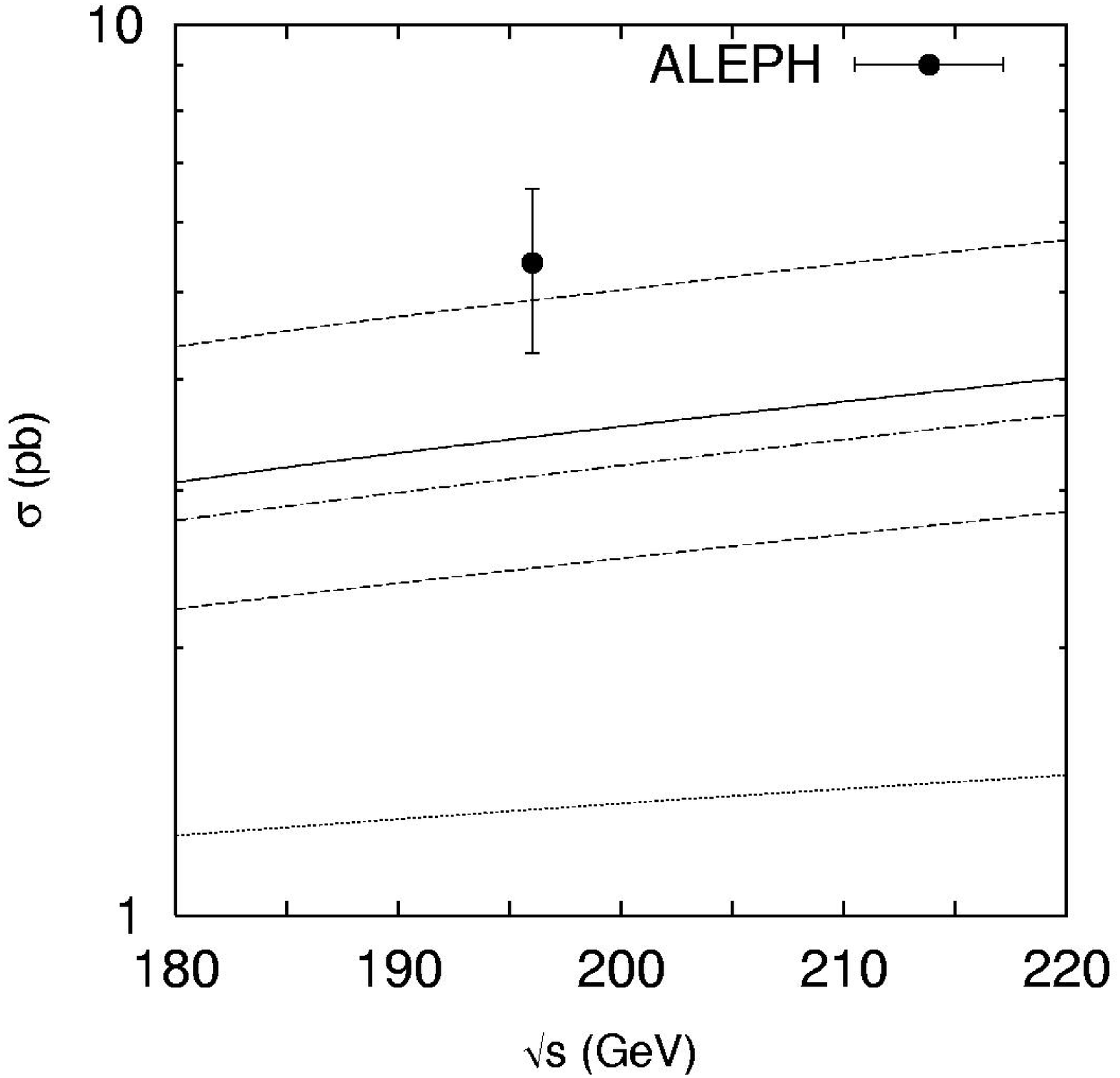, width = 20cm}
\caption{The beauty total cross section $\sigma(e^+ e^- \to e^+ e^- b\bar b \, X)$
  as a function of the $e^+ e^-$ center-of-mass energy $\sqrt s$. 
  The solid and dash-dotted lines correspond to the results obtained with the 
  CCFM and KMR unintegrated gluon densities, respectively. The upper and lower dashed 
  lines correspond to the CCFM gluon density with variation in $b$-quark mass and scale 
  as it was described in text. Separately shown is the contribution from the direct 
  production mechanism (dotted line). 
  The experimental data are from ALEPH~[15].}
\end{center}
\label{fig1}
\end{figure}

\newpage

\begin{figure}
\begin{center}
\epsfig{figure=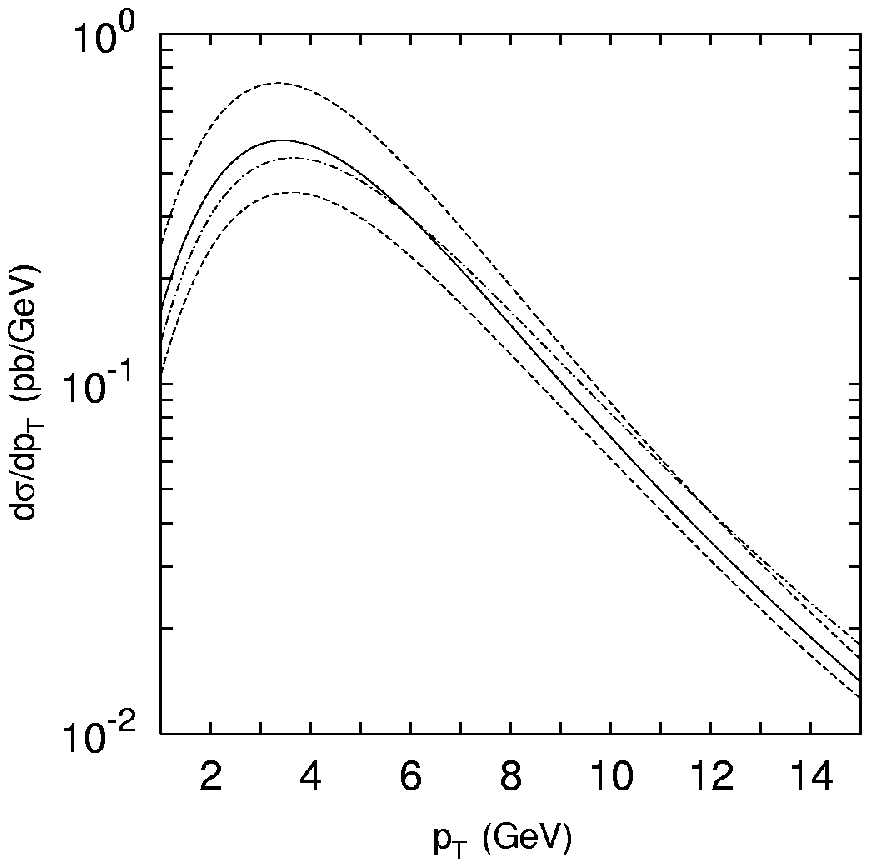, width = 20cm}
\caption{The differential beauty cross section $d\sigma/dp_T$ for 
the process $e^+ e^- \to e^+ e^- b\bar b \, X$ at $|\eta| < 2$ and $\sqrt s = 196$~GeV. 
Notation of curves is the same as in Fig.~1.}
\end{center}
\label{fig2}
\end{figure}

\newpage

\begin{figure}
\begin{center}
\epsfig{figure=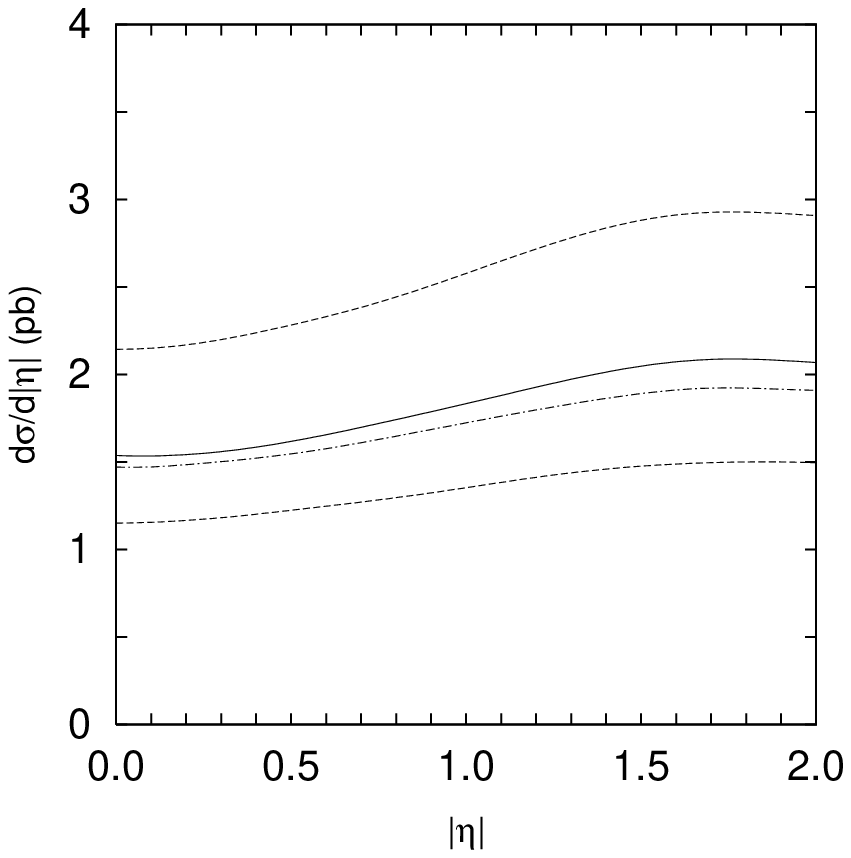, width = 20cm}
\caption{The differential beauty cross section $d\sigma/d|\eta|$ for 
the process $e^+ e^- \to e^+ e^- b\bar b \, X$ at $p_T < 20$~GeV and $\sqrt s = 196$~GeV.
Notation of curves is the same as in Fig.~1.}
\end{center}
\label{fig3}
\end{figure}

\newpage

\begin{figure}
\begin{center}
\epsfig{figure=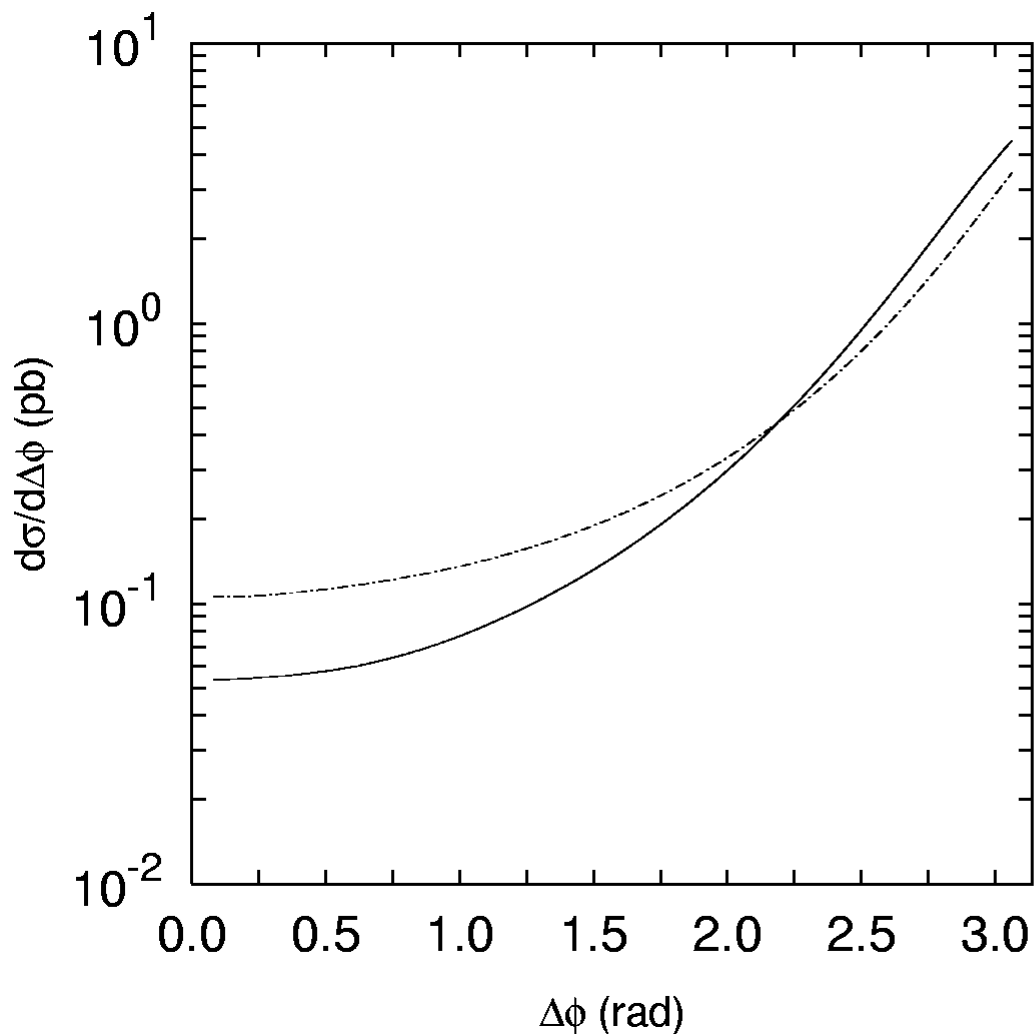, width = 20cm}
\caption{The differential beauty cross section $d\sigma/\Delta \phi$ for 
the process $e^+ e^- \to e^+ e^- b\bar b \, X$ at $\sqrt s = 196$~GeV.
Notation of curves is the same as in Fig.~1.}
\end{center}
\label{fig4}
\end{figure}

\end{document}